\documentclass[oldversion]{aa} 
\usepackage{graphicx} 
\usepackage{subfigure} 
\usepackage{natbib} 
\usepackage{epsfig} 
\usepackage{ulem} 
      \def\new#1 {{\textbf #1 }} 
      \def\cut#1 {\sout{#1} } 
 
\bibpunct{(}{)}{;}{a}{}{,} 
 
\def\folio{\ifnum\pageno=1\nopagenumbers\else\number\pageno\fi}

\def\lax    {\ifmmode{_<\atop^{\sim}}\else{${_<\atop^{\sim}}$}\fi}
\def\gax    {\ifmmode{_>\atop^{\sim}}\else{${_>\atop^{\sim}}$}\fi}
\newbox\grsign      \setbox\grsign=\hbox{$>$}
\newdimen\grdimen   \grdimen=\ht\grsign
\newbox\simgreatbox \setbox\simgreatbox=\hbox{\raise.5ex\hbox{$>$}\llap
                        {\lower.5ex\hbox{$\sim$}}}\ht1=\grdimen\dp1=0pt
\newbox\simlessbox  \setbox\simlessbox =\hbox{\raise.5ex\hbox{$<$}\llap
                        {\lower.5ex\hbox{$\sim$}}}\ht2=\grdimen\dp2=0pt

\newbox\grsign \setbox\grsign=\hbox{$>$} \newdimen\grdimen \grdimen=\ht\grsign
\newbox\laxbox \newbox\gaxbox
\setbox\gaxbox=\hbox{\raise.5ex\hbox{$>$}\llap
     {\lower.5ex\hbox{$\sim$}}}\ht1=\grdimen\dp1=0pt
\setbox\laxbox=\hbox{\raise.5ex\hbox{$<$}\llap
     {\lower.5ex\hbox{$\sim$}}}\ht2=\grdimen\dp2=0pt
\def\gax{\mathrel{\copy\gaxbox}}
\def\lax{\mathrel{\copy\laxbox}}
\def\boxit#1    {\vbox{\hrule\hbox{\vrule\kern3pt
                  \vbox{\kern3pt#1\kern3pt}\kern3pt\vrule}\hrule}}
\def\h      {\ifmmode{^{\rm h}}\else{$^{\rm h}$}\fi}
\def\m      {\ifmmode{^{\rm m}}\else{$^{\rm m}$}\fi}
\def\s      {\ifmmode{^{\rm s}}\else{$^{\rm s}$}\fi}
\def\am     {\ifmmode {\rlap.}$\,$'$\,$\! \else ${\rlap.}$\,$'$\,$\!$\fi}
\def\decam     {\ifmmode {\rlap.}$\,$'$\,$\! \else ${\rlap.}$\,$'$\,$\!$\fi}
\def\decas    {\ifmmode{{\rlap.}{''}}\else{${\rlap.}{''}$}\fi}
\def\mum     {\ifmmode{\mu{\rm m}}\else{$\mu{\rm m}$}\fi}
\def\decam {\rlap . {}'}          
\def\am {\rlap . {}'}          
\def\s      {\ifmmode{^{\rm s}}\else{$^{\rm s}$}\fi}
\def\decdeg {\rlap . {}^\circ}     
\def\deg      {\ifmmode{^{\circ}}\else{$^{\circ}$}\fi}
\def\as     {\ifmmode {\rlap.}$\,$''$\,$\! \else ${\rlap.}$\,$''$\,$\!$\fi}
\def\decsec  {\ifmmode {\rlap.}$\,$^{\rm s}$\,$\! \else ${\rlap.}$\,$^{\rm s}$\,$\!$\fi}
\def\decs  {\ifmmode {\rlap.}$\,$^{s}$\,$\! \else ${\rlap.}$\,$^{s}$\,$\!$\fi}

\def\kms    {\ifmmode{{\rm km~s}^{-1}}\else{km~s$^{-1}$}\fi}

\def\Lsun   {$L_{\odot}$}

\def\Mspy   {\ifmmode {M_{\odot} {\rm yr}^{-1}} \else $M_{\odot}$~yr$^{-1}$\fi}
\def\Mdot   {\ifmmode {\dot M} \else $\dot M$\fi}
\def\mhd    {\ifmmode {n_{{\rm H}_2}} \else $n_{{\rm H}_2}$\fi}
\def\mhcd   {\ifmmode {N_{{\rm H}_2}} \else $N_{{\rm H}_2}$\fi}

\def\El      {\ifmmode{E_{\ell}}\else{$E_{\ell}$}\fi}
\def\beam    {\ifmmode{\theta_{\rm B}}\else{$\theta_{\rm B}$}\fi}
\def\Jyb   {\ifmmode {{\rm Jy~beam}^{-1}} \else{Jy~beam$^{-1}$}\fi}
\def\mjyb   {\ifmmode {{\rm mJy~beam}^{-1}} \else{mJy~beam$^{-1}$}\fi}
\def\mujyb   {\ifmmode {\mu{\rm Jy~beam}^{-1}} \else{$\mu$Jy~beam$^{-1}$}\fi}
\def\Trot   {\ifmmode{T_{\rm rot}}\else$T_{\rm rot}$\fi}

\def\Teff   {\ifmmode{T_{\rm eff}}\else$T_{\rm eff}$\fi}

\def\ITRS   {\ifmmode{\smallint {\rm T}_{R}^{*}dv}\else{$\smallint
{\rm T}_{R}^{*}dv$}\fi}
\def\ITRS   {\ifmmode{\smallint {\rm T}_{R}^{*}dv}\else{$\smallint
{\rm T}_{R}^{*}dv$}\fi}
\def\ITAS   {\ifmmode{\smallint {\rm T}_{A}^{*}dv}\else{$\smallint
{\rm T}_{A}^{*}dv$}\fi}

\def\lefttitle#1  {\noindent \hangindent=18.0pt \hangafter=1 {#1} \par}
\def\vol#1  {{\bf {#1}{\rm,}\ }}
\font\tenssb=cmssbx10
\font\tenbf=cmbx10
\font\sevenbf=cmbx8
\font\fivebf=cmbx6
\def\unetdemi    {\smallskipamount=6pt plus2pt minus2pt
                  \medskipamount=12pt plus4pt minus4pt
                  \bigskipamount=24pt plus8pt minus8pt
                  \normalbaselineskip=16pt plus0pt minus0pt
                  \normallineskip=2pt
                  \normallineskiplimit=0pt
                  \jot=6pt
                  {\def\smallskip {\vskip\smallskipamount}}
                  {\def\medskip   {\vskip\medskipamount}}
                  {\def\bigskip   {\vskip\bigskipamount}}
                  {\setbox\strutbox=\hbox{\vrule
                    height17.0pt depth7.0pt width 0pt}}
                  \parskip 12.0pt
                  \normalbaselines}
\def\smallerspace {\smallskipamount=3pt plus0pt minus0pt
                  \medskipamount=6pt plus0pt minus0pt
                  \bigskipamount=10.5pt plus0pt minus0pt
                  \normalbaselineskip=10.5pt plus0pt minus0pt
                  \normallineskip=1pt
                  \normallineskiplimit=0pt
                  \jot=3pt
                  {\def\smallskip {\vskip\smallskipamount}}
                  {\def\medskip   {\vskip\medskipamount}}
                  {\def\bigskip   {\vskip\bigskipamount}}
                  {\setbox\strutbox=\hbox{\vrule
                    height8.5pt depth3.5pt width 0pt}}
                  \parskip 0pt
                  \normalbaselines}
\def\memospace    {\smallskipamount=4pt plus1pt minus1pt
                  \medskipamount=6pt plus2pt minus2pt
                  \bigskipamount=14pt plus6pt minus6pt
                  \normalbaselineskip=14pt plus0pt minus0pt
                  \normallineskip=1pt
                  \normallineskiplimit=0pt
                  \jot=4pt
                  {\def\smallskip {\vskip\smallskipamount}}
                  {\def\medskip   {\vskip\medskipamount}}
                  {\def\bigskip   {\vskip\bigskipamount}}
                  {\setbox\strutbox=\hbox{\vrule
                    height17.0pt depth7.0pt width 0pt}}
                  \parskip 2.0pt
                  \normalbaselines}
\def\memowidespace    {\smallskipamount=5pt plus1pt minus1pt
                  \medskipamount=7.5pt plus2pt minus2pt
                  \bigskipamount=17.5pt plus6pt minus6pt
                  \normalbaselineskip=17.0pt plus0pt minus0pt
                  \normallineskip=1.25pt
                  \normallineskiplimit=0pt
                  \jot=5pt
                  {\def\smallskip {\vskip\smallskipamount}}
                  {\def\medskip   {\vskip\medskipamount}}
                  {\def\bigskip   {\vskip\bigskipamount}}
                  {\setbox\strutbox=\hbox{\vrule
                    height21.25pt depth8.75pt width 0pt}}
                  \parskip 2.5pt
                  \normalbaselines}
\def\masj {mas~yr$^{-1}$} 

\begin{document} 
 
\title{The size, luminosity and motion of the extreme carbon star
IRC+10216 (CW Leonis)}                                                  
\author{K. M. Menten 
\inst{1} 
\and 
M. J. Reid 
\inst{2} 
\and 
T. Kami{\' n}ski 
\inst{1} 
\and 
M. J. Claussen 
\inst{3} 
} 
 
\offprints{K. M. Menten} 
 
\institute{Max-Planck-Institut f\"ur Radioastronomie, 
Auf dem H\"ugel 69, D-53121 Bonn, Germany 
\email{kmenten, kaminski@mpifr-bonn.mpg.de} 
\and 
Harvard-Smithsonian Center for Astrophysics, 
60 Garden Street/MS42, Cambridge MA 02138, USA 
\email{reid@cfa.harvard.edu} 
\and 
National Radio Astronomy Observatory, 
Array Operations Center, P.O. Box O, 
Socorro, 
NM 87801, USA 
\email{mclausse@nrao.edu} 
} 
\date{Received / Accepted} 
\titlerunning{Size, luminosity and motion of IRC+10216} 
\authorrunning{Menten et al.}

\abstract{Very Large Array observations of the extreme carbon star
IRC+10216 at 7 mm wavelength with 40 milli-arcsecond resolution
resolve the object's radio emission, which forms an almost round
uniform disk  of 83 milli arcseconds diameter, corresponding to 11
AU (for an assumed distance of 130 pc). We find a brightness temperature
of 1630 K for the radio photosphere. Since the emission is optically
thick, we can directly estimate IRC+10216's average luminosity, which is
8600 \Lsun. 
This value is in excellent agreement with 
what is predicted from the period-luminosity relation for carbon-rich Miras.
Assuming an effective temperature of 2750 K for IRC+10216, it implies an optical
photospheric diameter of 3.8 AU. 
Our precise determination of IRC+10216's
proper motion fits the picture presented by far-ultraviolet and far-infrared 
wavelength observations of its interaction region with the
interstellar medium (its ``astrosphere''): the star moves roughly
in the direction expected from the morphology of the termination
shock and its astrotail structures.        
Calculation of its three dimensional velocity   
and an analysis of the kinematics of its surrounding interstellar medium (ISM) suggest an appreciable relative velocity of 42 \kms, which is about half
the value discussed in recent studies. This suggests a lower (time-averaged) mass loss 
rate and/or a higher ISM density than previously assumed.
 
\keywords{ISM: molecules  -- Stars: circumstellar matter}} 
 
\maketitle 
 
\section{\label{intro}Introduction} 


The carbon-rich evolved star IRC+10216 (also known 
as CW Leonis) is one of the most prominent and  best-studied 
near-infrared (NIR) sources in the sky \citep{Becklin1969}. IRC+10216
is probably a                                                           
typical carbon star \citep{Herbig1970,Miller1970} near the end of
its lifetime on the asymptotic giant branch (AGB), which is             
characterized by extreme mass-loss \citep{HabingOlofsson2003}.  This
and its proximity,                                                      
make IRC+10216 a unique object of interest, allowing studies
that would be                                                           
very difficult or impossible for any other source of its kind. Estimated
distances, $D$, inferred from modeling the CO emission in its
envelope, range from 110--150 pc \citep{Crosas1997, Groenewegen1998}; we
shall adopt a median value of 130~pc.
 
IRC+10216's high mass-loss rate of $2\cdot10^{-5}$ \Mspy\ \citep[][scaled
to 130 pc]{Crosas1997}                                                  
results in a dense 
circumstellar envelope (CSE) whose exceedingly rich 
chemistry can be easily studied at infrared through submillimeter
to radio wavelengths                                                    
\citep[see, e.g. ][]{Cernicharo1996, Cernicharo2000, Patel2011}.
The innermost part of the envelope, i.e., within $\sim$50\,AU 
or 20 stellar radii, and the star itself is hardly observable 
at visual wavelengths, due to the high extinction of the dust that
is abundantly produced in this region.                                  
In the infrared (IR) regime, however,  IRC+10216 presents a complex and 
dynamical picture.  High spatial resolution imaging obtained with 
speckle interferometry and/or adaptive optics shows several distinct 
features on subarcsecond scales that vary over time-scales of years, 
not only in position but also in luminosity  \citep{Osterbart2000,
Weigelt2002,Leao2006}.                                                  
Combined data taken in the near/mid-IR $H$, $K$, $L$, $M$, and $N$ bands, 
together with far-IR data, have been used to 
find a self-consistent model of the star and its envelope 
\citep{Menshchikov2001}. 
 
All modeling efforts are 
hampered, however, by the inability to tell which (or if any) 
of the observed compact features actually corresponds to the stellar
photosphere, with different approaches yielding widely differing conclusions
\citep[see above references and][]{Tuthill2005}. 
Moreover,  adaptive optics $H$-band 
imaging polarimetry by \citet{Murakawa2005} 
implies that the position of the 
illumination source (the central star) is different from any of the 
previously 
postulated positions. 
Needless to say, all radiative transfer modeling efforts of this
keystone                                                                
envelope, sophisticated as they may be \citep[see, e.g. ][]{Ivezic1996,
Groenewegen1997, Menshchikov2001}                                       
severely suffer from the uncertainty of not knowing 
the position from which the luminosity originates. 
 
Clearly, a direct detection of the star that could unambiguously determine the                                       
stellar position would be superior to any of the above 
indirect methods. One of the objectives of the present study is to
determine IRC+10216's position at short radio wavelengths (7 mm)
with an accuracy of a few milli arcseconds (mas) using the NRAO\footnote{The
National Radio Astronomy Observatory (NRAO) is operated by Associated
Universities, Inc., under a cooperative agreement with the National
Science Foundation.} Very Large Array.                                  
In a related project we seek to achieve absolute 
infrared astrometry to match our accurate radio astrometry. Since
the radio position marks the bona fide location of the star, it will thus eventually
be possible to unambiguously determine its IR counterpart.              
 
Radio emission from IRC+10216 has been studied  at wavelengths of
2 and 1.5 cm \citep{Sahai1989, Drake1991}.                              
\citet{Menten2006} presented Very Large Array observations
at 3.6, 2, and 1.3 cm, which revealed an unresolved source $<95$ mas 
and
established that the emission is 
optically thick, i.e., its spectral
index, $\alpha$, is $\approx 2$; where flux density, $S$, is $\propto
\nu^\alpha$. This allows us to address a second goal of the present
higher resolution ($\sim40$ mas) study, which is to determine the
size and shape of IRC+10216's radio emission distribution and, using
its implied brightness temperature, its luminosity.                     
 
Our highly accurate position, combined with published data allow a
high quality determination of IRC+10216's proper motion on the sky. This
is an important quantity, given recent ultraviolet and far-infrared
(FIR) observations of the interaction region between the star's expanding
circumstellar envelope and the ambient interstellar medium.             
 
This paper has the following structure: In \S\ref{observations} we
give an account of our 
VLA observations. The determination
of IRC+10216's size, accurate position, and 
proper motion are described in \S\ref{results}.  The nature
of its radio emission, the luminosity it implies, and its
motion through the interstellar medium are discussed in \S\ref{discussion}.
 
\section{\label{observations}Observations and data processing} 
\subsection{VLA observations} 
Our VLA observations took place on 2006 February 26 (JD 2453793)
with 23 antennas in operation. We used the maximum                      
bandwidth setting provided by the VLA correlator. 
This comprises 2 intermediate frequency (IF) bands, each with 43
MHz effective                                                           
bandwidth. We recorded, both, right and left circular polarization 
in each IF. One IF was centered at 43.3149 GHz, the
other at a 50 MHz higher frequency. 
 
In order to ensure optimal calibration for our 
high frequency data, we employed the ``fast switching'' technique first 
described by \citet{Lim1998}.  Over 7.5 h, we switched between
IRC+10216 and one of the two nearby calibrators, J0943+170 or J0954+177.  
We used a cycle with 60 seconds on IRC+10216 and 40 seconds on one of the
calibrators, which we repeated for 1--1.5 h segments. 
In order to check the effectiveness of our observing mode, between segments 
we switched between the two calibrator sources for $\sim10$ min.
 
We improved the VLA antennas' pointing accuracy by
updating pointing offsets once an hour, by observing J0954+177 in
``interferometer pointing mode''.  For higher sensitivity these 
pointing scans were taken in X-band (8.4 GHz). The corrections to the 
pointing model were applied real time.        
 
While processing the data we found that the position of J0943+170 
as listed in the VLA calibrator database was insufficiently accurate,
given the precision warranted by our data.  We adopted
better positions for both calibrator sources from the VLBI Global 
Solution 2008b Astro Catalog maintained at the NASA Goddard Space Flight
Center\footnote{http://lacerta.gsfc.nasa.gov/vlbi/solutions/}; see
\citet{Petrov2008}.             
We used the task CLCOR within NRAO's Astronomical Image 
Processing System (AIPS) to correct our data for the improved
positions.                                                               
The adopted calibrator positions, as well as the position of IRC+10216
determined as described in \S\ref{results}, are listed in Table \ref{pos}.

\begin{table*}[!t] 
 \caption{\label{pos}IRC+10216 and calibrator sources} 
\begin{center} 
\begin{tabular}[!t]{llllllll} 
Source            & $\alpha({\rm J}2000)$ &$\delta({\rm J}2000)$
& ${\rm d}x$ &${\rm d}y$ & $S$\\                                        
                      &                    &                    
&($^{\circ}$)&($^{\circ}$)&(Jy)\\                                       
\noalign{\smallskip} 
\hline 
IRC+10216 (CW Leo)&$09^{\rm h}47^{\rm m}57\decsec4255(6)$&$+13^{\circ}16'43\as815(10)$& -- & -- 
& 0.013\\
J0943+170                  & 09 43 17.22396(2)         &+17 02 18.9628(12)&$-1\decdeg1$  & $+3\decdeg8$ & 
0.106 \\
J0954+177                  & 09 54 56.82362(5)         &+17 43 31.2222(2)&$+1\decdeg7$ & $+4\decdeg4$ & 
0.234 \\
3C286 (J1331+3030)& 13 31 08.28806(2)         &+30 30 32.9592(5)&       --          &  --&1.455\\
\noalign{\smallskip} 
 \hline 
 \noalign{\smallskip} 
\end{tabular} 
\end{center} 

 \textbf{Notes.} Measured position for IRC+10216 (epoch 2006.16) and the adopted positions for
the calibrator sources are given in columns two and three in J2000 coordinates. 
Calibrator positions are taken
from the VLBI Global Solution 2008b Astro Catalogue available on
the NASA Goddard Spaceflight  Center Geodetic VLBI group's website;
see also \citet{Petrov2008}.                                            
Numbers in parentheses give the error in the last quoted digit(s). 
For the complex-gain calibrators J0943+170 and J0954+177, column four
lists their angular separations from IRC+10216. The separation between
these two calibrators is $2\decdeg9$.  Column six lists the measured flux
densities based on the assumed value for 3C286.
We estimate the flux density scale is accurate to
$\pm5$\%.                                        
\end{table*}

\subsection{\label{cali}Calibration and initial $uv$-plane and image
analysis}                                                               
Calibration and image processing was performed with 
AIPS. 
While loading the data into AIPS (using the task FILLM), 
corrections for the elevation-dependent gain curve and
atmospheric transmission were applied.                                  
Visual inspection of the measured visibilities 
revealed a minimal amount of flawed data. 
The absolute flux density, $S$, scale was established by an observation of 3C286,
which has $S = 1.455$ Jy at                                             
43.317 GHz. Since this source is slightly 
resolved at our observing frequency, we used a clean-component model
downloaded from the VLA website to determine amplitude and phase  
solutions using the AIPS task CALIB. 
Then we used CALIB to determine amplitude and phase solutions for
J0954+177 and J0943+170,                                                
for which we bootstrapped flux densities of 0.234 and 0.106 Jy, respectively,
by comparing with the solutions found for 3C286.                        
Using CLCAL, we applied the amplitude and phase corrections to the
IRC+10216 data.                                                         
 
Using IMAGR we produced an image of IRC+10216 and found the star
offset in right ascension and declination direction, $(\Delta x, \Delta y)$, by 
 $(+0\as641, +0\as157)$ relative
to our phase-center position. To enable subsequent $uv$-domain         
analysis, we used UVFIX to shift the visibility data to the center of 
the \textit{uv-}plane. We then used IMAGR to produce images of the shifted     
\textit{uv-}dataset, both with uniform and natural weighting (see \S\ref{imageanalysis})
 
\section{\label{results}Results} 
%
\subsection{\label{imageanalysis}Imaging of IRC+10216} 
 
\begin{table*}[!t] 
 \caption{\label{runs}Model fits to IRC+10216's 43.3 GHz brightness
distribution}                                                           
\begin{tabular}[!t]{lllllllll} 
&\multicolumn{3}{l}{Restoring beam}&\multicolumn{3}{l}{Deconvolved
source}\\                                                               
M&$\theta_{\rm B,maj}$ & $\theta_{\rm B,min}$ & P.A.(B) & $\theta_{\rm
S,maj}$ & $\theta_{\rm S,min}$ & P.A.(S) & $S_{\rm p}$  &$\int S{\rm
d}\Omega$\\ 
&(mas)  & (mas)  &($^\circ$, E of N)  & (mas)   &   (mas)  & ($^\circ$,
E of N) & (mJy~beam$^{-1}$) &  (mJy)\\ 
\noalign{\smallskip} 
\hline 
\multicolumn{9}{l}{Imaging with natural weighting -- Gaussian model}\\ 
J&61    &   49    &   $-6.3$ &  $60\pm2$     &$54\pm2$  &  $+34\pm17$&$6.22\pm0.07$ &  $13.1\pm0.2$\\ 
$13.1\pm0.2$&2630\\
\multicolumn{9}{l}{Imaging with uniform weighting -- Gaussian model}\\ 
J&44    &   39    &   $-6.8$ &  $63\pm2$     &$57\pm2$  &  $+26^{+16}_{-22}$& $4.30\pm0.09$ & 
$13.2\pm0.4$\\ 
$13.3\pm0.3$&2450\\
\multicolumn{9}{l}{Visibility versus $uv$-distance -- Gaussian modeling}\\
O&--      & --        &--             &  $61\pm1$     &$55\pm$3&$+31\pm13$      & --            & $12.8\pm0.2$\\ 
L&--      &--         &--             &  $56\pm2$     &-- &--                   &--             & $12.6\pm0.3$\\  
\multicolumn{9}{l}{Visibility versus $uv$-distance -- uniform disk modeling}\\
O&--     &   --      &--              &  $87\pm2$     &$80\pm1$ &$+22\pm5$      & --            & $12.2\pm0.1$\\
L&--     & --        & --             &  $83\pm1$     &--       &--             &--             &  $12.1\pm0.2$\\
 \noalign{\smallskip} 
 \hline 
 \noalign{\smallskip} 
\end{tabular} 
 
\textbf{Notes.} The first four lines give results of Gaussian fitting, line 1 and
2 to a naturally weighted image and a uniformly weighted image, line
3 and 4 to visibility domain data. Lines 5 and 6 give the results
for fits with a uniform disk model. The first column denotes the         
method (M) used: J for JMFIT and O for OMFIT (AIPS tasks). L stand
for our independent least-squares program. For the first two lines,
the second to forth columns give the FWHM major and minor axes and
the position angle of the restoring beam determined by IMAGR, while
the fifth to seventh columns list the same quantities for IRC+10216's
brightness distribution and the eighth column gives fitted peak brightness.
For all fits, the ninth gives the integrated flux density (the "zero
spacing" flux).
JMFIT and OMFIT fitted elliptical brightness
models, whereas our least squares program used a circular model.
\end{table*} 
 
\begin{figure}[h] 
\begin{center} 
\includegraphics[width=0.4\textwidth,angle=0]{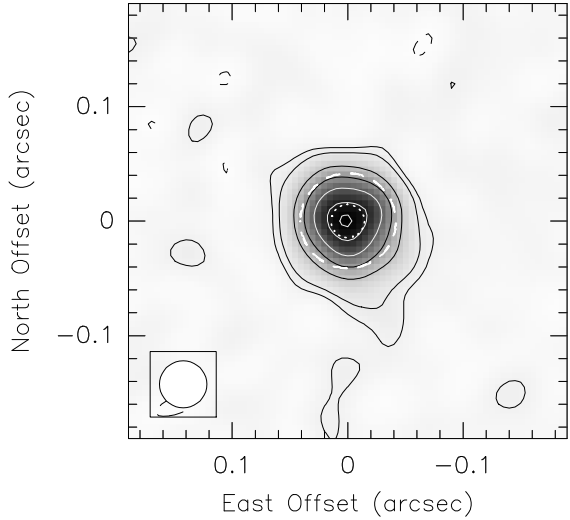} 
\end{center} 
\caption{Image of IRC+10216 at 43.3 GHz (6.9 mm). The brightness
distribution is shown in grey scale. Contours give $-5$, 5, 10, 30,
50, 70, 90, and 99 percent of the peak brightness, which is 4.1 mJy~beam$^{-1}$.
The rms noise level in the image, $92~\mu$Jy, corresponds to 2.2\%\
of that peak  value. The 41 mas FWHM symmetric restoring beam is
represented in the lower left corner. The 83 mas diameter dashed
circle gives the size of IRC+10216's radio photosphere derived by fitting a uniform disk
brightness distribution to our VLA data. The dotted circle of diameter 29 mas
represents the diameter of the star's photosphere derived from the luminosity we determine and assuming an \cut{efficient} \new{effective} temperature of 2850 K.
Note that the major axes of Mars' and Jupiter's orbits (3.0 and 10.4 AU, respectively) 
would have angular
diameters of 23 and 80 mas at a distance of 130 pc.}                            
\label{fig1} 
\end{figure} 
 
\begin{figure}[h] 
\begin{center} 
\includegraphics[width=0.4\textwidth,angle=0]{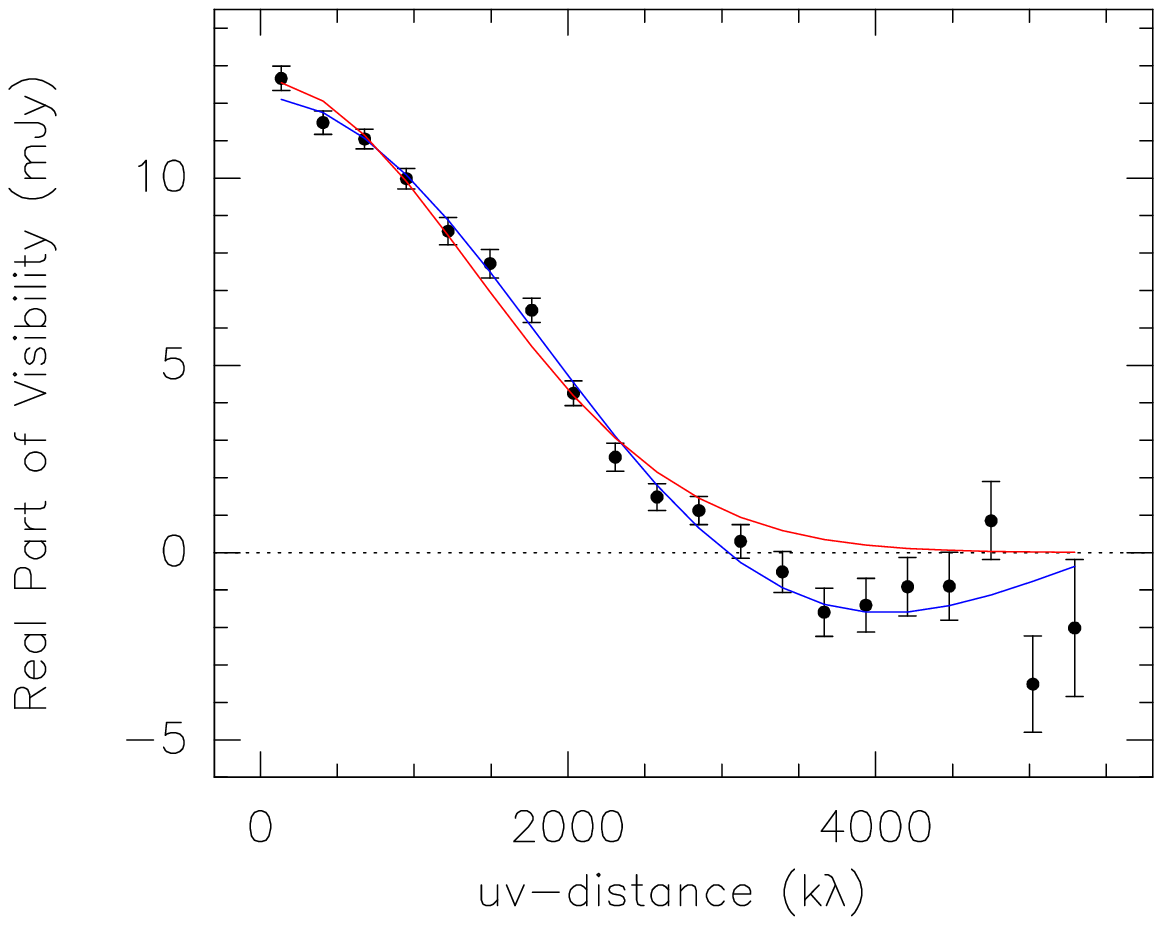} 
\end{center} 
\caption{Fringe visibility vs. baseline length for IRC+10216
at 43.2 GHz (6.9 mm) on 2006 February 26.                               
The red line represents the Fourier transform of a Gaussian brightness distribution with
an FWHM of 56 mas and the blue line that of a uniform disk of diameter
83 mas (see text and Table \ref{runs}).                                 
} 
\label{fig2} 
\end{figure} 
 
\subsubsection{\label{size}The size and brightness temperature of the radio emission}   
Table \ref{runs} shows the result of fits to IRC+10216's observed
brightness distribution at 43.3 GHz.  Fits were made to the image (using
JMFIT) as well as to its corresponding $uv$-plane data (using OMFIT).
We solved for the peak brightness and 
source size using elliptical models. Our results, represented in
Table \ref{runs} and Fig. \ref{fig2}, indicate an almost circular
source. Given this, we also performed fits with our
independently developed least-squares fitting program, which assumes circular symmetry.
This has the advantage that it delivers useful statistical quantities.
An inspection of Fig. \ref{fig2} indicates that the
uniform disk model (blue line) matches the data much better than
the Gaussian (red line).  This is verified by a comparison of the reduced
$\chi^2$ values, which are 1.35 for the disk and 3.07 for the Gaussian
model.
The best fit uniform disk has a diameter, $d$, of $83\pm1$ mas,
corresponding to $10.8\pm0.1$ AU ($D = 130$~pc) and a brightness temperature, $T_{\rm B}$, of $1635\pm82$ K.          
The uncertainty in $T_B$ is dominated by the 5\%\  estimated uncertainty of our absolute 
calibration.           
 
\subsection{\label{pospms}Position and proper motion of IRC+10216} 
\subsubsection{Position} 
Using JMFIT, we determined the best-fit position for IRC+10216 listed
in Table \ref{pos}.                                                     
The formal precision returned by JMFIT in the eastward, $x$, and northward, $y$, directions are 0.4 mas, much smaller than the values quoted in the table. 
To arrive at the listed, more realistic values for these quantities, we performed
several tests.  First, we only allowed phase and amplitude corrections determined for 
J0943+170 (in the following the ``weak'' calibrator) to be used 
to calibrate the IRC+10216 data. About half of the data, those taken
while switching between IRC+10216 and the ''strong'' calibrator, J0954+177, were discarded. 
Second, vice verse, we calibrated IRC+10216 only with the strong calibrator, 
discarding the data taken while switching with the weak calibrator. 
We then imaged both calibrated datasets and again used JMFIT to determine
the resultant positions.  We find position centroids with offsets, 
($\Delta x, \Delta y$), relative to the position
in Table \ref{pos} of ($-2.8$ mas, $-6.4$ mas) and (+2.8 mas, +5.9 mas), respectively. 
IRC+10216's best fit flux density determined from both of these sub-datasets
agreed with that determined from the whole dataset to within 2\%. 
 
To get a further  estimate of the accuracy of our position determination,
we calibrated the $uv$-data of the weak calibrator with the 
phase and amplitude corrections determined for the strong calibrator. 
Imaging of the former retrieved J09433+170's position with an offset of 
($\Delta x, \Delta y$) = ($-1.3$ mas, $-3.4$ mas) relative to its
nominal value quoted above.                                             
Calibrating, vice versa, strong with weak calibrator retrieved J0954+177's
position at an offset of  (+6.5 mas, +9.1 mas).                         
In this exercise, both calibrator sources appeared unresolved and
JMFIT returned formal upper limits of 7 and 10 mas for sizes of the
weak and the strong calibrator, respectively. The flux densities of         
both were found to within a few percent of the bootstrapped value
quoted above.                                                           
We note that the weak and strong calibrator databases only contained
20\% and 14\% of the number of visibilities of the IRC+10216 database.    
 
As listed in Table \ref{pos}, the separation on the sky between
IRC+10216 and J0943+170 and J0954+177 is $3\decdeg9$ and $5\decdeg9$,
respectively, and thus are 1.4 and 3.1 times larger than the $2\decdeg9$
arc between the two calibrators. Thus, it seems prudent to            
scale the position errors resulting from the calibrator-calibrator
experiment by the ratio of the $x$ and $y$ differences. This would result
in an  absolute position uncertainty of $\approx 3$ mas in each coordinate.
We conservatively increase this to 10 mas, because of  the fact that our calibrators are both 
offset in declination and astrometric accuracy is expected to be worse for target-calibrator separations that are mostly in 
the north-south direction.  
 
\subsubsection{\label{promo}Proper motion} 
\begin{figure}[h] 
\begin{center} 
\includegraphics[width=0.4\textwidth,angle=0]{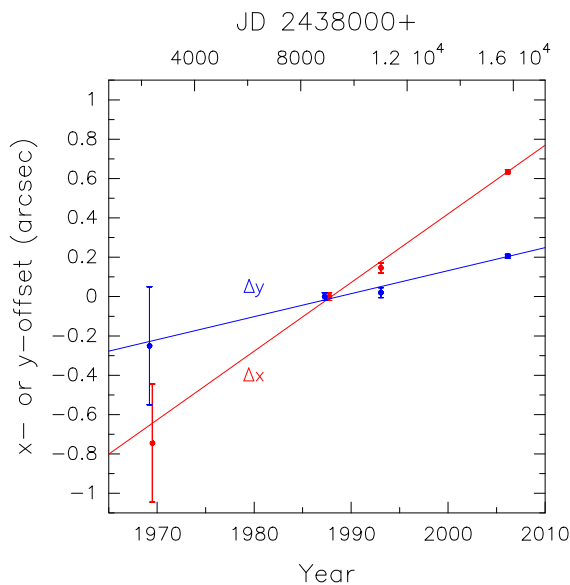} 
\end{center} 
\caption{Proper motion of IRC+10216 determined for epochs 1969.36
\citep{Becklin1969}, 1987.51 and 1993.07 \citep[both][]{Menten2006}, 
and 2006.16 (this paper). The red and blue dots mark the position 
offset relative to the epoch 1987.51 position in the eastward (x) 
and northward (y) directions, respectively. The
error bars for the 2006 position are comparable to the symbol size.
Lines represents least-squares fits to the data. 
For the first two epochs the times have been slightly offset for clarity.}                                       
\label{pm} 
\end{figure}

\citet{Becklin1969}, in the very first publication on IRC+10216, placed an upper limit 
of 30 mas~yr$^{-1}$ on its proper motion. These authors compared a position measured
on a plate from 1969 with the position from an E-plate
of the \textit{Palomar Sky Survey} taken in 1954.   \citet{Menten2006}
used positions determined from VLA observations in 1987 and 1993 (with uncertainties of $\approx 25$ 
mas) 
to determine IRC+10216's proper motion in $x$- and
$y$-direction, ($\mu_x, \mu_y$), of $ (+26\pm6, +4\pm6$) \masj. Combining
these data with our more accurate position from 2006 (Table 1), we
calculate a proper motion of ($+35\pm1, +12\pm1$) \masj\ (see Fig. \ref{pm}).
For a distance of 130 pc, these angular motions correspond to linear speeds
of 21.6~\kms\ eastward and 7.4~\kms\ northward.     
This proper motion is a heliocentric value and needs to be corrected for the solar motion. 
The resulting values are  36.3~\kms\ eastward and 13.2~\kms\ northward, implying a speed 
in the plane of the sky of 38.6 \kms\ along a  position 
angle  of  70$^\circ$ (E of N).
We note that the motion we determine is in the opposite
direction of that implied by the binary model for IRC+10216, which
\citet{Guelin1993} propose to explain an offset they
find between the star and the centroid of the C$_4$H emission 
imaged with the IRAM Plateau de Bure interferometer.                    
 
 \citet{Roeser2010} combine data from the optical USNO-B1.0 all sky
survey and the Two Micron All Sky Survey (2MASS) to calculate astrometric
parameters for common sources, IRC+10216 amongst them. In the resulting
PPMXL catalog, IRC+10216's proper motion is listed as ($\mu_x, \mu_y$)
=  ($+28\pm6, -8\pm6$) \masj.   Their eastward motion
is in agreement with ours, while their northward motion is
discrepant by $\approx 3\sigma$. 
Given the complex structure of IRC+10216's innermost CSE and the                                                   
significant variation of its morphology with wavelength, combined
with the uncertainty of whether or not the stellar photosphere is observed
at optical and IR wavelengths (see Sect. \ref{intro}), one
might question the reliability of the PPMXL proper motion determination.    
 
\section{\label{discussion}Discussion} 
\subsection{Properties and nature of the radio emission} 
\subsubsection{Size and luminosity} 
Our measured radio brightness temperature of 1660 K is definitely
lower than IRC+10216's optical effective temperature, $T_*$.  After
critically evaluating models of AGB stars and a wide range of data
(their Table 4) \citet{Menshchikov2001} conclude that $T_*$ lies between
between 2500 and 2800 K.   Thus, our radio
size and temperature are in line with what \citet{ReidMenten1997,
ReidMenten2007} find for the ``radio photosphere,'' which surrounds the 
optical photospheres of M-type (oxygen-rich) AGB stars.  They find that
the radius of the radio photosphere is about twice the
stellar radius and the temperatures is about $1/\sqrt2$ lower than
at the stellar surface.  Modeling the ionization balance and opacity
sources  in these regions (mainly H$^-$ free-free interactions),
\citet{ReidMenten1997} predict a spectral index, $\alpha_{\rm RP}$,
for the radio emission of 1.86.                                         
 
From measurements taken between 8 and 680 GHz, \citet{Menten2006}
determined the radio-to-submillimeter spectral index of IRC+10216's  
radio emission to be $1.96\pm0.04$, 
which is steeper than the value 
predicted for radio photospheres by \citet{ReidMenten1997} and consistent 
with optically-thick blackbody emission in the Rayleigh-Jeans regime
($\alpha_{\rm BB} \equiv 2$).  
Since we are observing emission consistent with a blackbody, our measured diameter, 
$d$ and brightness temperature, $T_{\rm B}$, directly deliver IRC+10216's
bolometric luminosity, $L_*$, via the Stefan-Boltzmann law: $L_*
= \pi d^2 \sigma T_{\rm B}^4$, where $\sigma$ is the Stefan-Boltzmann constant. 
With $T_{\rm B} = 1635\pm82$ K and $d$= $10.8\pm0.1$~AU (83 mas at $D = 130$~pc), 
we calculate $L_* = 3.3\cdot10^{37}~{\rm erg~s}^{-1}$, or  $8640\pm430$~\Lsun.             
The last uncertainty \textit{does not} include a contribution from the distance uncertainty.

Although IRC+10216's bolometric luminosity varies by a factor of 2.5 over a light cycle,   
model calculations predict that its radio flux varies by no more than $\pm10$\%\ between minimum and 
maximum \citep[][and pers. comm.]{Menshchikov2001}. This is in excellent agreement with IRC+10216's 
$850~\mu$m light curve derived from data taken with the James-Clerk-Maxwell-Telescope\footnote
{http://www.jach.hawaii.edu/JCMT/continuum/} and within their 
errors consistent with the radio data discussed by \citet{Menten2006}. Given this, a luminosity derived 
from radio data at any time is close to the star's average luminosity.  
Our estimate gains in reliability since we were able to determine IRC+10216's long-wavelength (infrared) 
phase, $\phi_{\rm IR}$\footnote{Conventionally, the phase, $\phi$, quoted for an AGB star is the time, 
measured as a 
fraction of its period, that has passed since optical maximum. Due to molecule formation, for most AGB stars, the IR 
maximum lags $\approx 0.2$~ periods behind the optical maximum. This is well-established for oxygen-rich Miras, but less so for carbon-rich ones \citep{Smith2006}.}, at the time of our VLA 
observations. As described in Appendix \ref{phasedet},  we 
acquired data to determine a  reliable value for $\phi_{\rm IR}$, of 0.75.  This means the luminosity we have determined is very close to the average value indeed.
 
The (average) luminosity we determine, 8640 \Lsun\ (for $D = 130$ pc), is  lower than
the value of 14000 \Lsun\ that \citet{Groenewegen1998} find
consistent with their multi-transition modeling of the CO emission
from IRC+10216's CSE, while \citet{Crosas1997}  and \citet{deBeck2012} 
find 11000 and 8500 \Lsun, respectively, in similar analyses (all for $D = 130$ pc).  

If we assume $T_{\rm eff} = 2750$~K for IRC+10216's effective temperature, i.e., the
median value considered by \citet{Menshchikov2001} for their modeling, we calculate a diameter
of 3.8 AU for the star's photosphere. This corresponds to $\approx 1.3$ times the major axis of Mars' 
orbit.

The derived luminosity is comparable to the value one derives from
the revised period--luminosity relation for carbon-rich Miras established
by \citet{GroenewegenWhitelock1996}. 
For our best fit period, 630 d (see Appendix \ref{phasedet}),
this relation predicts a luminosity of  9830 \Lsun.
Finally, we note that, remarkably, 
\citet{Becklin1969} in the very first publication on IRC+10216 derive $10^{31}$~W for its luminosity (assuming 
$D = 200$~pc), which corresponds to 11000 \Lsun\ for $D = 130$~pc.
 
\subsubsection{Shape of the radio photosphere} 
Performing VLA observations with angular resolution and wavelength similar to ours, 
\citet{ReidMenten2007} determined the shapes and sizes of the radio photospheres of 
three M-type AGB stars ($o$ Ceti, R Leo and W Hya); the latter were all close
to 6 AU.  While $o$ Ceti's radio photosphere has, like IRC+10216's,
an almost circular shape, those of R Leo and W Hya appear to be
significantly elongated. Formally, we find a flattening, $e~(\equiv
(a-b)/a)$ of $0.08\pm0.02$ for IRC+10216 (where $a$ and $b$ are the major
and minor axis sizes, respectively).  From Table 2 of \citet{ReidMenten1997}
we calculate $e = 0.07\pm0.13, 0.36\pm0.14$ and $0.33\pm0.14$ for $o$ Ceti, 
R Leo and W Hya, respectively
 
IRC+10216's radio photosphere is almost perfectly round, comparable
to that of the much lower mass-loss star $o$ Ceti. Given its high
mass-loss rate, IRC+10216 has been conjectured to be on the verge of
developing into a protoplanetary nebula (PPN) and asymmetries in
its circumstellar material have been interpreted as signs for the
onset of bipolarity \citep{Osterbart2000}.  Here we emphasize that
the radio emission is consistent with an extended photospheric origin
and does not show the characteristics of a developing ultracompact
HII region as observed in the more evolved PPN CRL 618. For this
object, which has a $\sim 30$ times higher mass loss rate than IRC+10216
\citep{Young1992}, VLA observations show an enhanced radio flux associated
with an extended ionized region that is elongated along the PPN's
bipolar outflow axis \citep{Kwok1984,MartinPintado1993}

\subsection{Motion through the interstellar medium} 
Recent observations at ultraviolet, FIR and submillimeter wavelengths
provide evidence for an interaction of IRC+10216's outflowing envelope
and the ambient interstellar medium (ISM). Knowledge of the star's
motion is an important parameter in any modeling of this phenomenon.    
 
\begin{figure}[h] 
\begin{center} 
\includegraphics[height=\columnwidth,angle=-90]{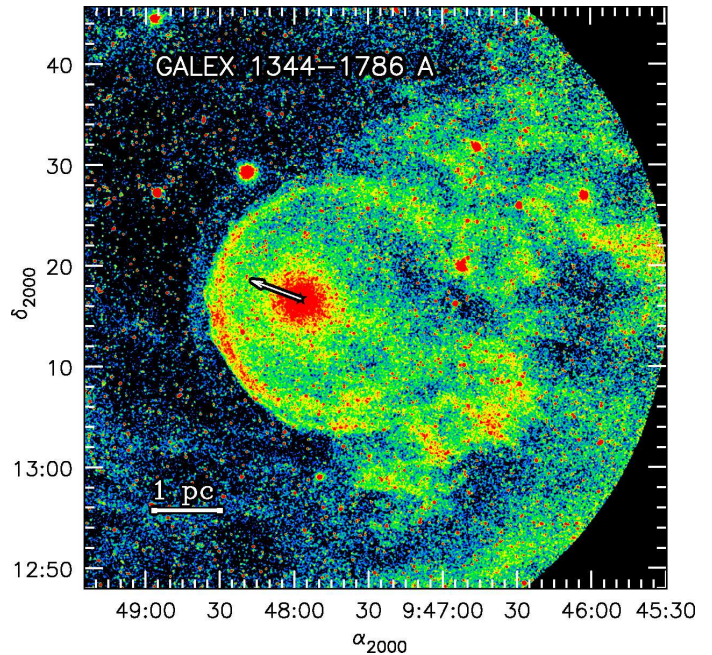} 
\end{center} 
\caption{IRC+10216 as seen by GALEX in 2008 in the FUV band \citep[see][]{Sahai2010}.
The image has been smoothed from the original resolution with a Gaussian kernel with a FWHM 
of 5 pixels. The star symbol indicates the position of IRC+10216 and the 
arrow represents its proper motion (corrected for the solar motion) for 5,000 yr. } 
\label{fig4} 
\end{figure}

\citet{Sahai2010} present near and far ultraviolet (NUV and FUV)
imagery obtained with the Galaxy Evolution Explorer (GALEX) of an area of
$\approx 1$~degree diameter centered on IRC+10216, which shows the
object's ``astrosphere'' (Fig. \ref{fig4}).  It exhibits a textbook
picture of phenomena expected from the interaction of a circumstellar
outflow with the ambient interstellar medium: an astrosheath, 
astropause, a termination shock, and an astrotail \citep[see Fig.
2 of ][for an illustration of these terms]{Ueta2008}. An arc-like
segment of the termination shock is also detected. The arrow in (Fig. \ref{fig4}) 
shows  the direction of IRC+10216's proper motion (corrected for the 
Solar motion) and the distance traveled 
by the star in 5\,000~yr.                 
 
Remarkably, the termination shock also manifests itself as a dust
shell that has been imaged with \textit{Herschel} at 160 and 250~$\mu$m
with the Photodetector Array Camera and Spectrometer (PACS) and
the SPectral and Photometric Imaging REceiver (SPIRE) \citep{Ladjal2010}.
Moreover, \citet{Decin2011}, using PACS at 70, 100, and 160~$\mu$m,
find  multiple concentric dust shells with the outermost one reaching
almost to the termination shock.                                        
 
For modeling such interactions between stellar winds and the interstellar
medium (ISM), one critical parameter is the velocity of the star relative
to the ISM, \textit{v}$_*$, which we can estimate from the proper
motion we have derived in Sect. \ref{promo}; see Fig. \ref{fig4}.
In the FUV image the astrosheath forms an incomplete edge brightened
shell (the termination shock) around the stellar position. However,
this shell is broken up toward the WSW--SW, in which direction (most
of) the structures making up the astrotail extend beyond its boundaries.
We note that this is the direction expected for the astrotail for
the proper motion vector we determine, which has a position angle
of 70$\deg$ (E of N).                                                 
 
Theoretical studies \citep{vanBuren1988, Wilkin1996} have yielded
a formula for the scale, $l_1$, at which the ram pressure of the
freely flowing wind from a moving star equals that of the interstellar
medium.  This is measured as the distance between the termination
shock and the star along the astropause's symmetry axis and is a function
of mass loss rate, $\dot{M}_*$, terminal wind velocity, \textit{v}$_w$,
mean molecular mass per atom, $\bar{\mu}_{\rm H}$, ISM number density,
$n_{\rm ISM}$, and \textit{v}$_*$. Solving for the stellar velocity,
we have                                                                 
$$ 
 v_* \propto (\dot{M} ~\textit{v}_w /  \bar{\mu}~n_{\rm ISM})^{1/2}/l_1 
$$ 
Using the $l_1$ determined by \citet{Sahai2010}, adopting 
$\dot{M}_*$ of $2\cdot10^{-5}$~\Mspy\ (see \ref{intro}), and 
\textit{v}$_w$ = 14~km~s$^{-1}$), and assuming a low density atomic ISM 
($\bar{\mu}_{\rm H}$ and $n = 1.33$ and $1$~cm$^{-3}$), respectively, \citet{Sahai2010} derive a stellar
velocity of 91~\kms. \citet{Ladjal2010} obtain a slightly higher
value of $107 \pm 9$~\kms\  (for $n_{\rm ISM} = 1$~cm$^{-3}$).  
These values assume that the astropause's symmetry axis lies in the
sky plane.

To compare the above velocities with our proper motion, we must transform our heliocentric value into
the galactic reference (LSR) frame and correct for the solar motion. Following the instructions of \citet{Johnson1987AJ}, we 
transform our heliocentric proper motions and radial velocity, 
$v_{\rm Hel}$\footnote{We calculate this heliocentric velocity for IRC+10216
assuming an LSR radial velocity, $v_{\rm LSR}$, of $-26$~\kms, using the solar
motion vector $(+10,+5,+7)$ \kms, derived from Hipparchos data by \citet{DehnenBinney1998}. Note that our value of  $v_{\rm LSR}$ differs from the often quoted $-22$~\kms\ \citep[as, e.g., 
given in the catalog compiled by][]{Loup1993}. The latter value is derived from low-$J$ CO line profiles that over estimate the LSR velocity  because of self absorption in their blue wings.}, 
of $-19$~km~s$^{-1}$\
to a velocity vector in the galactic frame of reference, $(U, V, W)_{\rm Hel}$, which has to be corrected for 
the solar motion, to, finally, obtain $(U, V, W)_{\rm LSR}$ in the LSR frame.  Here, the components $U$, $V$, and $W$ are positive in the directions of the  Galactic center, Galactic rotation and the North Galactic pole.
%
If we do this transformation, we obtain 
$(U, V, W)_{\rm LSR}$ = (+34,+23,+10) \kms\ for IRC+10216, which has an absolute value of 42 \kms\  with an estimated uncertainty of less than 10 \kms. 

We now have determined the three dimensional motion vector for IRC+10126 in the LSR frame. What we 
do 
not know is the motion of the ISM in that frame. For the  radial component of this motion, 
we can get some insight into this question by examining 
our knowledge of the ISM around the star. As discussed in Appendix \ref{radvel},  the radial component of IRC+10216's motion  is between $13$ and $26$~ km\,s$^{-1}$ lower than that to its ambient ISM.

Having no other constraints, we assume
that this ISM is partaking in Galactic rotation 
with no peculiar velocity contribution, i.e., $(U,V,W)^{\rm ISM}_{\rm LSR}$ = (0,0,0) \kms. Then, our best estimate value of the motion of IRC+10216
relative to its 
ambient ISM, 42 \kms,  is a factor of $\approx 2$--3 smaller than the values quoted by \citet{Ladjal2010} and \citet{Sahai2010}.


We note that several assumptions made by the above authors could
have led to significant uncertainties in their estimate of $v_*$.
For one, mass loss rates of the order of IRC+10216's, $2\cdot10^{-5}$ \Mspy, 
(or even higher)
are characteristic with the latest stages of a star's AGB lifetime
\citep{Volk2000, Perinotto2004}. It is thus very  unlikely, that
IRC+10216's mass loss rate had its present very high value over the
several times $10^4$ yr time scale it took the astrosphere to be established.
In fact, concentric shells in CSEs as observed around IRC+10216 by
\citet{Decin2011}, have been interpreted as evidence for variable
mass loss and are  possibly the result of multiple thermal pulses
\citep{Olofsson1990, Zijlstra1992}.                                     
 
Second, the ambient ISM's density, for which \citet{Sahai2010} assume
1~cm$^{-3}$, is poorly constrained. In the case of Mira  ($o$ Ceti),
which also excites FUV emission while moving through the ISM \citep{Martin2007},
modeling suggests a 50 times smaller value \citep{Wareing2007, Ueta2008}.
Excitation of molecular hydrogen (H$_2$) by hot electrons is considered
the best candidate mechanism for the FUV emission from
IRC+10216's astrosphere \citep{Martin2007, Sahai2010}. The presence
of molecular hydrogen would imply that IRC+10216 is moving within
a diffuse molecular cloud. Diffuse \textit{molecular} clouds, however,
have hydrogen number densities of  100--500~cm$^{-1}$ \citep{Snow2006},
significantly larger than the 1 cm$^{-1}$ assumed by \citet{Sahai2010}
(and $1.33 < \bar{\mu}_{\rm H} < 2.33$).                                
As a caveat, we point out that Mira's FUV emission distribution
looks distinctly different from IRC+10216's. It has a very long,
much more turbulent tail with an elongated cometary shape and, while
it has a pronounced bow shock, it lacks the thin ring making up more
than a semi circle that marks the boundary of IRC+10216's astropause.
For the latter, hopefully, further modeling will result in a more
meaningful estimate of the ISM density and other quantities.            
At present, the above facts, plus the unknown inclination angle,
prohibit a meaningful estimate of $v_*$ from the UV observations.
Vice versa, we note that even if we assumed that $v_*$ were equal
to our estimated speed (42~\kms) 
the uncertainties addressed above would make any conclusions based
on the UV emission of IRC+10216 and its envelope, e.g., on the duration
of the AGB phase, extremely uncertain.

\section{Conclusions} 
Using the VLA we have imaged the radio emission from the archetypical
high mass loss carbon-rich AGB star IRC+10216. We obtain a precise
measure of the size of its radio emission (83 mas) and its brightness
temperature (1660 K). Since the emission is consistent with arising
from a black body, these quantities, together with the assumed 
distance of 130 pc, determine the average stellar luminosity to be 8640
\Lsun. 
This value is in excellent agreement with 
what is predicted from the period-luminosity relation for carbon Miras.
 
Our position determination, with  $\approx10$ milli arcsecond accuracy, together
with older data, allow a determination of IRC+10216' s proper motion.
After correcting for the solar motion, we find that the star moves with 
a velocity of 39 \kms\ in the plane of the sky
in east-northeastern direction (position angle 70$^\circ$) 
and has a velocity of 42 \kms\  with respect to the LSR.
The direction of this
motion is roughly  consistent with FUV and FIR images that show an
extended astrosphere around  IRC+10216, which has a broken shell and tail-like features in the
opposite direction. Our calculation of the star's three dimensional velocity   
and an analysis of the kinematics of its surrounding ISM suggest a 
lower relative velocity than derived in recent studies. This suggests a lower (time-averaged) mass loss rate and/or a higher ISM density than previously assumed.

\begin{acknowledgements}
We thank the group of V. I. Shenavrin for making their IR data publicly
available and O. Taranova for explanatory remarks concerning the
light curves. We are grateful to Alexander Men`shchikov for providing information on his model calculations, Bo Zhang and Andreas Brunthaler for independent model fitting, Bryan Butler for help with OMFIT and Martin Groenewegen for a clarifying remark.
\end{acknowledgements}

\begin{appendix}
\section{\label{phasedet}Phase determination for the epoch of the VLA observations}

\begin{figure}[h] 
\begin{center} 
\includegraphics[angle=270,width=0.4\textwidth]{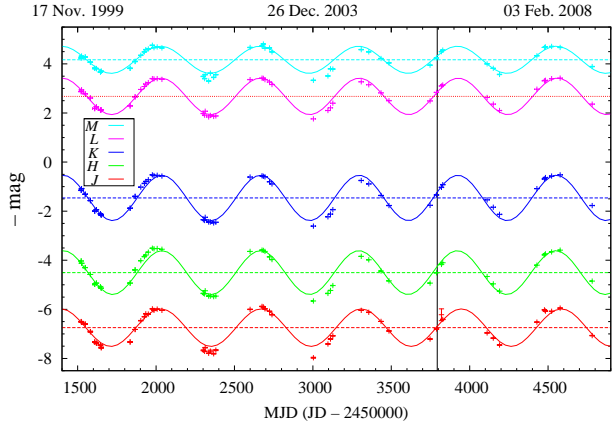} 
\end{center}
\caption{Bottom to top: $J, H, K, L, $ and $M$ infrared light curves of IRC+10216 (from \citet{IRphot}) and 
the our cosine fits to the variability curves. The vertical line represent the epoch of the VLA observations 
(2006 February 23).}                            
\label{lcurve}
\end{figure}

In order to derive the phase of IRC+10216's light curve at the epoch of our VLA observations (2006 
February 23), we used infrared photometry from \citet{IRphot}. This compilation consists of 310 
measurements in the $JHKLM$ bands obtained between 1999 December 10 and 2008 November 11 
(see Fig.~\ref{lcurve}). Each of the light curves was interpolated by a cosine fit using an implementation of 
the nonlinear least-squares Marquardt-Levenberg algorithm in gnuplot\footnote{http://www.gnuplot.info}. 
The period ($P$), the epoch of the chosen maximum (JD$_{\rm max}$), the average magnitude, and the 
amplitude of variations were treated as free parameters. For the $HKLM$ bands, this procedure resulted 
in periods between 628 and 632 days with typical uncertainty of 1 day, and the moment of maximum 
between 2454550 and 2454560 with typical uncertainty of 4 days. The $J$ light curve has slightly poorer 
quality data and the fit gave results slightly different than for the other bands ($P$=638.5 and JD$_{\rm 
max}$=2454585.0) and was discarded from the further analysis. The combined values for the $HKLM$ 
light curves, in terms of weighed-mean and rms, are $P$=630.0$\pm$2.9 days and JD$_{\rm max}
$=2454554.0$\pm$7.4. 
This is consistent with the range of values \citet{LeBertre1992} determined for $P$ from monitoring in different IR bands, 635--670 d (with an average of 649 d). His data were taken in the mid 1980s over a time range of duration less than half than that of our dataset.
We next used these values to compute the phase of our VLA observations, which 
gave $\phi$=0.79$\pm$0.06 (where $\phi$=0 corresponds to the maximum brightness). This result 
agrees very well with the value of $\phi$=0.72 computed from the stellar luminosity variations given in 
\citet{Menshchikov2001} (their Eq. 1), which was based on IR variability of the source between 1965 and 
1998. 

\section{\label{radvel}On the radial component of the motion of IRC+10216 relative to its local ISM}

\begin{figure}[t] \begin{center} 
\includegraphics[height=10cm,angle=270,width=0.3\textwidth]{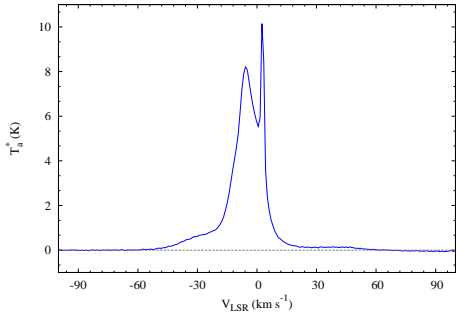} 
\end{center}
\caption{The profile of the 21 cm line of \ion{H}{I} averaged over a large region around IRC+10216 
chosen to represent the typical line of sight emission of neutral hydrogen within a few degrees from the 
star. Data extracted from the GALFA survey \citep{galfa}.}                            
\label{HIprofile} 
\end{figure}

Insight into the interstellar vicinity of IRC+10216 can be gained from \ion{H}{I} observations toward the 
source and around it. For this purpose we extracted a data cube from the GALFA survey \citep{galfa}, 
which covers the full astropause of IRC+10216 and a large region around it at a resolution of 4\arcmin. 
The \ion{H}{I} emission, although inhomogeneous on angular scales comparable to the size of the 
astropause, is well represented by the profile shown in Fig.\,\ref{HIprofile}. It is dominated by two 
components centered around --5 km\,s$^{-1}$ and +3 km\,s$^{-1}$ and with FWHM line widths of 11 and 
1.5 km\,s$^{-1}$, respectively. The broader component has very extended wings seen from --60 to 60 km
\,s$^{-1}$. At some positions around the astropause, a weaker component can be seen centered at about 
--25 km\,s$^{-1}$. There are no morphological arguments for any connection of the observed emission 
with the astrosphere of IRC+10216. The \ion{H}{I} data itself do not allow to distinguish which components 
can be physically collocated with the star, if any.

Much has been learned about the ISM toward IRC+10216 by studies of absorption lines seen 
in optical spectra of nearby field stars. \citet{Kendall2002} identified two stars at angular distances of 
37\arcsec\ and 137\arcsec\ from the position of IRC+10216, which are located behind its circumstellar 
envelope (at spectro/photometric distances of 0.5 kpc and 1.4 kpc). The resonance lines of \ion{Na}{I} 
and \ion{K}{I} show clear signatures of the CSE material but do not show any extra component that could 
be identified with the ISM. It cannot be ruled out, however, that some atomic absorption is hiding within 
the strong profiles of the circumstellar absorption. The same authors find the diffuse interstellar band 
(DIB) at 6284 \AA\ in the spectra of the background stars at $v_{\rm Hel} = $ $-2.9$~km\,s$^{-1}$ 
(corresponding to $v_{\rm LSR} = $ $-9.6$~km\,s$^{-1}$) which they interpret as an ISM feature. They find 
the same velocity component in the profile of the \ion{Na}{I} lines of a star which is seen outside the 
astrosphere of IRC+10216 (2\fdg5 away from the the stellar position) and is located at a distance of 55\,pc. 
This suggests that the DIB originates in a nearby, {\it foreground} cloud. In another approach to the same 
observational material, \citet{MaH2010} found absorption features of \ion{Ca}{II} centered at the expected 
velocity of the circumstellar material but broader than the \ion{K}{I} lines. One interpretation is that 
interstellar material is partially responsible for those features, but a fully circumstellar origin cannot be 
excluded and is actually suggested by the symmetric  broadening of the profiles.

These optical studies indicate that the \ion{H}{I} emission at positive LSR velocities most likely arises in 
far, background clouds, while the component at --10 km\,s$^{-1}$ is probably related to a foreground 
cloud. Given the monotonic character of Galactic rotation, it is very unlikely that gas at even more 
negative velocities is situated between these two mentioned components, so it is reasonable to conclude 
that also clouds at velocities lower than --10 km\,s$^{-1}$ are foreground objects. The monotonicity of Galactic rotation and our results for IRC+10216's motion discussed above also imply that 
the star is moving though a medium which has an LSR velocity somewhere between $-10$~km\,s$^
{-1}$ and 3 km\,s$^{-1}$. Given that IRC+10216's LSR velocity is  $-26$~ km\,s$^{-1}$, the radial component of its motion  is between $16$ and $29$~ km\,s$^{-1}$ smaller than that to its ambient ISM.

\end{appendix}



\end{document}